\documentclass[usenatbib]{mn2e}
\usepackage{epsfig} \usepackage{natbib} \usepackage{amssymb}
\usepackage{amsmath}

\newcommand{\lya}{Ly$\alpha$}

\def\simgt{\mathrel{\spose{\lower 3pt\hbox{$\sim$}} \raise
2.0pt\hbox{$>$}}} \def\simlt{\mathrel{\spose{\lower
3pt\hbox{$\sim$}}\raise 2.0pt\hbox{$<$}}}

\title[The EUV spectral index of $z\sim6$ quasars]{Near-zone sizes and the rest frame extreme ultra-violet spectral index of the highest redshift quasars}

\author[Wyithe J.S.B. \& Bolton J.S.]{J. Stuart B. Wyithe and James
S. Bolton
\\ School of Physics,
University of Melbourne, Parkville, Victoria 3010, Australia\\ 
Email:
swyithe@physics.unimelb.edu.au}

\date{Draft Version}
\pagerange{\pageref{firstpage}--\pageref{lastpage}} \pubyear{2006}

\def\LaTeX{L\kern-.36em\raise.3ex\hbox{a}\kern-.15em
    T\kern-.1667em\lower.7ex\hbox{E}\kern-.125emX}

\begin{document}

\label{firstpage}

\maketitle

\begin{abstract} 
The discovery of quasars with redshifts higher than six has prompted a great deal of discussion in the literature regarding the role of quasars, both as sources of reionization, and as probes of the ionization state of the IGM. However the extreme ultra-violet (EUV) spectral index cannot be measured directly for high redshift quasars owing to absorption at frequencies above the Lyman limit, and as a result, studies of the impact of quasars on the intergalactic medium during reionization must assume a spectral energy distribution in the extreme ultra-violet based on observations at low redshift, $z\la1$. In this paper we use regions of high Ly$\alpha$ transmission (near-zones) around the highest redshift quasars to measure the quasar EUV spectral index at $z\sim6$.  
We jointly fit the available observations for variation of near-zone size with both redshift and luminosity, 
and propose that the observed relation provides evidence for an EUV spectral index that varies with absolute magnitude in the high redshift quasar sample, becoming softer at  higher luminosity. Using a large suite of detailed numerical simulations we find that the typical value of spectral index for a luminous quasar at $z\sim6$ is constrained to be $\alpha=1.3^{+0.4}_{-0.3}$ for a specific luminosity of the form $L_\nu\propto\nu^{-\alpha}$. We find the scatter in spectral index among individual quasars to be in the range $\Delta \alpha\sim0.75-1.25$. These values are in agreement with direct observations at low redshift, and indicate that there has been no significant evolution in the EUV spectral index of quasars over 90\% of cosmic time.
\end{abstract}
    

\section{{INTRODUCTION}}
\label{introduction}

The discovery of distant quasars has allowed detailed absorption
studies of the state of the high redshift intergalactic medium (IGM)
at a time when the universe was less than a billion years old
\citep[][]{fan2006,willott2007,willott2009}. Beyond $z\sim6$ several
of these quasars show a complete Gunn-Peterson trough in their spectra
blueward of the Ly$\alpha$ line \citep[][]{white2003}. However, the
spectra of these distant quasars also show enhanced Ly$\alpha$
transmission in the region surrounding the quasar, implying the
presence of either an HII region \citep[][]{cen2000}, or of a
proximity zone.  The interpretation of these transmission regions has
been a matter of some debate. Different arguments in favour of a
rapidly evolving IGM at $z>6$ are based on the properties of the
putative HII regions inferred around the highest redshift quasars
\citep[][]{wyithe2004,wyithe2005,mesinger2004,kramer2009}. On the
other hand, \citet[][]{bolton2007}, \citet[][]{maselli2007} and
\citet[][]{lidz2007} have demonstrated that the features in high
redshift quasar spectra blueward of the Ly$\alpha$ line could also
be produced by a classical proximity zone. In this case, 
the spectra provide no evidence for a rapidly evolving
IGM. Importantly, these, and other analyses rely on an assumed value
for the EUV spectral index to convert from the observed
luminosity (redward of rest-frame Ly$\alpha$) to an ionizing flux,
which is the quantity of importance for studies of reionization. As a
further example, a value for the EUV spectral index must be assumed
to estimate the quasar contribution to the reionization of hydrogen
\citep[e.g.][]{srbinovsky2007}.

The spectral energy distributions of luminous quasars are thought to
show little evolution out to high redshift \citep[][]{fanrev2006}. For
example, broad emission line ratios at $z\sim5$ have similar values to
those observed at low redshift
\citep[][]{haman1993,dietrich2003}. Moreover, optical and infrared
spectroscopy of some $z\sim6$ quasars has indicated a lack of
evolution in the optical-UV spectral properties redward of Ly$\alpha$
\citep[][]{pentericci2003,vandenberk2001}. At higher energies, the
optical/IR -to-X-ray flux
ratios~\citep[e.g. ][]{brandt2002,strateva2005} and X-ray continuum
shapes~\citep[][]{vignali2003} show at most mild evolution from low
redshift. However \citet{jiang2010} have recently reported
hot-dust-free quasars at $z\sim6$ which have no counterparts in the
more local Universe. These quasars are thought to be at an early
evolutionary stage, and offer the first evidence that quasar
properties have evolved since the reionization era.

As yet, the spectral energy distribution blueward of the Lyman limit,
which is critical for studies of reionization, is not measured at
$z\sim6$ since the quasar spectra are subject to complete absorption.
In the future the EUV spectral index could be measured from 21cm
observations of the thickness of ionizing fronts of the HII regions
around high redshift quasars \citep[][]{kramer2008}. However in the
meantime the assumed value of the spectral index is usually
based on direct observations of the EUV spectrum of quasars at
$z\la1$, where they are not subject to absorption from the IGM.

In this paper we show that the extent of Ly$\alpha$ transmission
regions around high redshift quasars is sensitive to the EUV spectral
index, and so provides a probe of this otherwise hidden portion of the
spectrum. Our paper is presented in the following parts. We first
summarise quasar near-zones (\S~\ref{nearzone}), the observed
near-zone relation (\S~\ref{obs}), and our associated numerical
modelling (\S~\ref{NZ}). We then present constraints on the relation
between near-zone size, quasar redshift, and quasar luminosity
(\S~\ref{results}). In \S~\ref{HIIregion} and \S~\ref{EUV} we present
a discussion of the astrophysical interpretation of our results. As
part of our analysis we provide a revised estimate of the ionizing
background at $z=6$ and jointly constrain the ionizing background and
EUV spectral index. Some additional uncertainties in our modeling are
discussed in \S~\ref{other}, and our conclusions are presented in
\S~\ref{conclusion}.  In our numerical examples, we adopt the standard
set of cosmological parameters \citep[][]{komatsu2009}, with values of
$\Omega_{\rm b}=0.044$, $\Omega_{\rm m}=0.24$ and
$\Omega_{\Lambda}=0.76$ for the matter, baryon, and dark energy
fractional density respectively, and $h=0.72$, for the dimensionless
Hubble constant.

\section{Near-zones}

\label{nearzone}

In view of the difficulty of determining the physical conditions of
the quasar environment, as well as the observational challenges
associated with measuring the \lya~  transmission at high redshift,
\citet[][]{fan2006} defined a specific radius, hereafter referred to
as the {\em near-zone} radius, at which the normalised \lya~
transmission drops to 10 per cent after being smoothed to a resolution
of 20\AA. This value of 10 per cent is arbitrary, and is chosen to be
larger than the average Gunn-Peterson transmission in $z\simeq 6$
quasar spectra ($<4$ per cent). In order to make the measurement
unambiguous, the near-zone radius is defined at the point where the
transmission {\it first} drops below this level, even if it rises back
above 10\% at larger radii. The measured near-zone radius is therefore
dependent on the spectral resolution.

\begin{figure*}
\begin{center}
\includegraphics[width=13cm]{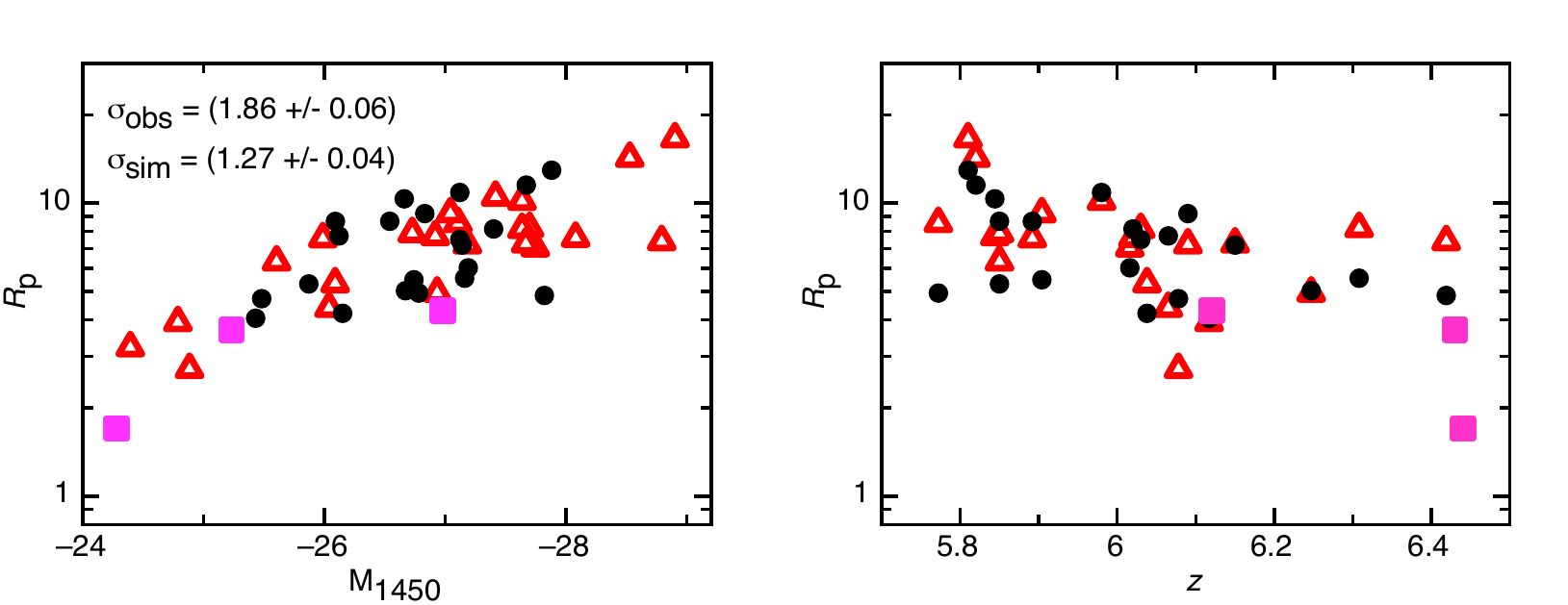}
\end{center}
\caption{\label{fig1}Measured sizes
  $R_{\rm p}$ plotted against $M_{\rm 1450}$ ({\em Left}) and $z$ ({\em Right}) respectively, for observed (black dots) and simulated (triangles) near-zones. Measured near-zone sizes are also plotted for CFHQS~J$1509-1749$, CFFQS~J$2329-0301$ and CFHQS~J$0210-0456$ (squares). The simulated  data has been computed at the best fit in $\alpha_0$ and $\alpha_1$ (see text for details).  }
\end{figure*}

\citet[][]{fan2006} found a striking relation between near-zone size and redshift, which they quantify using the expression
\begin{equation}
\label{nzrel}
R_{\rm p} = R_{\rm cor} +  A\times (z-6).
\end{equation}
Here the near-zone size, $R_{\rm cor}$, has been corrected for luminosity according to the relation 
\begin{equation}
R_{\rm corr} = R_{27} \times 10^{-0.4(27+M_{\rm 1450})/ B},
\end{equation}
where $B=3$ and $R_{27}$ is the mean size of a near-zone at $z=6$ around a quasar of $M_{\rm 1450}=-27$. The value of $ B=3$ \citep[which is roughly consistent with observations,][]{carilli2010} is motivated by the evolution of an HII region  \citep[][]{haiman2002} 
\begin{eqnarray}
\label{HII}
\nonumber
R_{\rm p} &=& 4.2 f_{\rm HI}^{-1/3}\left(\frac{\dot{N}}{1.9\times10^{57}\mbox{s}^{-1}}\right)^{1/3}\\
&&\hspace{10mm}\times\left(\frac{t}{10^7\mbox{yr}}\right)^{1/3}\left(\frac{1 + z}{7}\right)^{-1}\mbox{Mpc},
\end{eqnarray}
where $f_{\rm HI}$ is the fraction of hydrogen that is neutral,  $\dot{N}$ is the rate of ionizing photons produced by the quasar, and $t$ is the quasar age\footnote{This calculation assumes clumping and recombinations to be unimportant, that the density is at the cosmic mean over the large scales being considered, and that the quasar lifetime is much less than the age of the Universe.}. The size of the near-zones is found to increase by a factor of 2 over the redshift range from $6.4>z>5.7$, from which \citet[][]{fan2006} inferred the neutral fraction to increase by an order of magnitude over that time.

On the other hand, if the near-zone size is set by resonant absorption in an otherwise ionized IGM rather than by the boundary of an HII region, then its value is instead approximated by the expression
\begin{eqnarray}
\label{nz}
\nonumber
R_{\rm p} &\sim&\frac{3.1}{\Delta_{\rm lim}} \left(\frac{\tau_{\rm lim}}{2.3}\right)^{1/2}\left(\frac{T}{2\times10^4\mbox{K}}\right)^{0.35}\left(\frac{\dot{N}}{1.9\times10^{57}\mbox{s}^{-1}}\right)^{1/2}\\
&&\hspace{10mm}\times\left(\frac{\alpha^{-1}[\alpha+3]}{3}\right)^{-1/2}\left(\frac{1 + z}{7}\right)^{-9/4}\mbox{Mpc},
\end{eqnarray}
where $\Delta_{\rm lim}$ is the normalised baryon density in the transmitting regions as a fraction of the cosmic mean ($\tau_{\rm lim}=2.3$, corresponding to the 10 per cent transmission), and $T$ is the IGM temperature \citep[][]{bolton2007b}. The parameter $\alpha$ is the EUV spectral index. Equation~(\ref{nz}) differs from equation~(\ref{HII}) in three important ways. Firstly, if near-zone size is set by resonant absorption then it is not sensitive to the quasar lifetime. Secondly, resonant absorption results in a power-law index  describing the dependence of near-zone size on luminosity with a value of $ B=2$ rather than $ B=3$. Thirdly, the near-zone size is sensitive to $\alpha$, suggesting that near-zone size provides an opportunity to measure this unknown parameter.

The precise definition of near-zone radius suggested by \citet[][]{fan2006} allows for quantitative comparison with simulations. \citet[][]{wyithe2008} and \citet[][]{bolton2010} have modelled near-zones using hydro-dynamical simulations with radiative transfer, combined with a semi-analytic model for the evolving, density dependent ionizing background. The modelling in the latter work is able to produce both the amplitude and evolution of the near-zone sizes, as well as the statistics of absorption lines within the near-zone, and is based on quasars in a highly ionized IGM. In this regime it is equation~(\ref{nz}) with $B=2$, rather than equation~(\ref{HII}) with $B=3$ that is applicable. This indicates that a varying ionizing background could be responsible for the observed relation, providing an alternative scenario to a strongly evolving neutral fraction. Unfortunately, this scenario implies that the evolution of quasar near-zones cannot necessarily be used to infer an evolution in neutral fraction. 
Thus, although intriguing, it could be argued that the observed near-zone relation has not yet taught us anything concrete about reionization.

When discussing near-zones most authors have concentrated on the relation between size and redshift, with measured size corrected for luminosity dependence as part of this process. In most cases these studies used the scaling of near-zone in proportion to luminosity to the one third power \citep[e.g.][]{fan2006,carilli2010}. As described above this value is appropriate if the near-zone is expanding into a neutral IGM. On the other hand, a power of one half ($B=2$) is more appropriate if the near-zone corresponds to a radius where resonant absorption in an ionized IGM results in the 10 per cent transmission \citep[][]{bolton2007}. Thus the correct scaling between near-zone radius and quasar luminosity is uncertain. In this work we therefore treat the power-law index $B$ as a free parameter in a 2-dimensional relation describing near-zone radius as a function of quasar luminosity and redshift. We find that current near-zone data places strong constraints on the value of $ B$. Moreover, rather than focus on the interpretation of the parameter $A$ with respect to the end of hydrogen reionization, in this paper we concentrate instead on understanding the implications of the parameters $B$ and $R_{27}$ for the EUV spectral index blueward of the Lyman limit for $z\sim6$ quasars. We show that the near-zone relation offers a probe of the EUV spectrum of quasars at these early epochs, information that is otherwise concealed by the Lyman-limit absorption, and so has previously only been studied for quasars at redshifts of $z\la1$ \citep[][]{telfer2002}.

\section{Observed Near-Zone Sample}
\label{obs}

\citet{carilli2010} have assembled a sample of 25
$z\sim6$ quasars which have quality rest-frame UV spectra and redshift
measurements~\citep{fan2006,jiang2008}. Eight of the sample are detected~\citep{wang2010} in CO. Another nine have redshifts measured~\citep{kurk2007,jiang2007} from the Mg II line emission. For the other
eight objects, \citet{carilli2010} adopt redshifts from
the relevant discovery papers, which are mainly determined with the
Ly$\alpha$+NV lines~\citep{fan2006b}. The measurements are summarized
in Table 1 of \citet{carilli2010}, including the quasar
absolute AB magnitude at 1450 \AA~($M_{1450}$), quasar redshifts
($z$), and near-zone size ($R_{\rm p}$). Three sources in the original
sample \citep{wang2010} listed in Table 1 of \citet{carilli2010}, $J0353+0104$, $J1044-0125$, and $J1048+4637$,
are broad absorption line quasars \citep{jiang2008,fan2006b}, while
the source $J1335+3533$ has a lineless \citep{fan2006b} UV spectrum. We
exclude these quasars from our analysis
owing to the fundamentally different nature of their intrinsic
spectra~\citep{carilli2010,fan2006}. Thus the sample consists of 21 $z\sim6$ quasars which we use to analyse the near-zone relation.
Near-zone sizes are plotted in Figure~\ref{fig1} as a function of both absolute magnitude and redshift. Typical measurement
  errors~\citep{carilli2010} are 1.2 Mpc, 0.4 Mpc, and 0.1 Mpc in the
  estimated near-zone size ($R_{\rm p}$) introduced by UV, MgII, and
  CO-determined redshift uncertainties, respectively. There are clear
trends with both luminosity~\citep{carilli2010} and
redshift~\citep{fan2006,carilli2010}. 
  
  We note that quasar near-zone sizes have also been measured~\citep{willott2007,willott2010} for CFHQS~J$1509-1749$, CFFQS~J$2329-0301$ and CFHQS~J$0210-0456$. 
Following \citet{carilli2010} we do not include these near-zones, which are drawn from a different data set, in our analysis. However as may be seen from Figure~\ref{fig1}, these near-zones lie on the \citet[][]{carilli2010} correlation, and we have checked that  their addition  does not significantly alter our results.

\begin{figure*}
\begin{center}
\includegraphics[width=15cm]{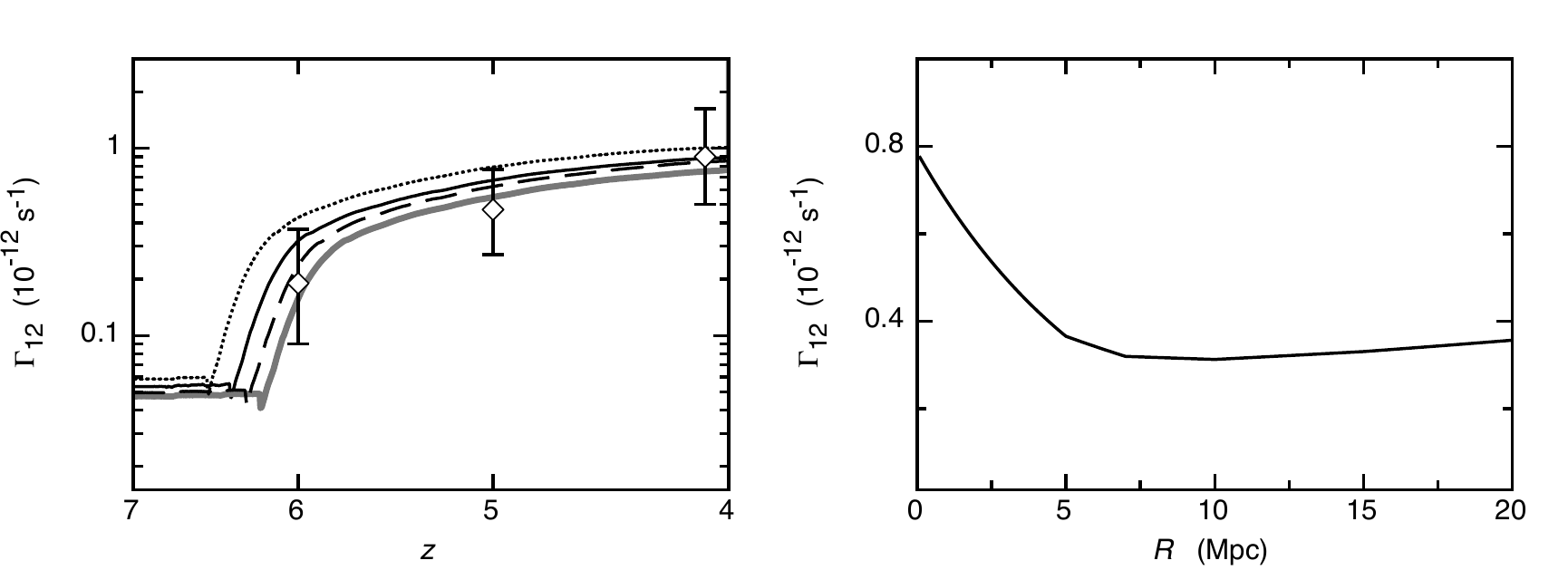}
\end{center}
\caption{\label{fig2}The fiducial model for the ionizing background. {\em Left:} The evolution of the
  ionizing background~\citep{wyithe2008}. The grey curve shows the
  background in the mean IGM, while the solid, dotted and dashed black
  curves show the mean and 1-sigma range of ionizing background in the
  biased regions within $5\,$Mpc of a $10^{13}\,$M$_\odot$ quasar host
  halo. The data points show our estimate of the ionizing background
  at $z=5$ and $z=6$ (error bars represent the 68\% range). The point
  at $z=4$ has been reproduced from \citet{bolton2007}. {\em Right:} The ionizing background as a
  function of radius in our fiducial model (for a quasar observed at
  $z=6$). The ionizing background has been computed as a function of
  proper time along the trajectory of a photon emitted by the quasar,
  rather than at the proper time of the quasar.}
  \end{figure*}

\section{Modelling of Near-Zones}
\label{NZ}

Our near-zone simulations combine a semi-analytical model for the
evolving, density dependent photo-ionization rate in the biased
regions surrounding quasars, with a radiative transfer implementation,
and realistic density distributions drawn from a high resolution
cosmological hydrodynamical simulation.  These simulations are
discussed in detail elsewhere ~\citep{bolton2010}, and their
description is not reproduced here.  However for completeness we show
the evolution of the background photo-ionization rate for our fiducial
simulations in the left hand panel of Figure~\ref{fig2}.  The grey
curve shows the photo-ionization rate in the mean IGM, while the
solid, dotted and dashed black curves show the mean and 1-sigma range
of the photo-ionization rate in the biased regions within $5\,$Mpc of
a $10^{13}\,$M$_\odot$ halo.  Importantly, we construct the absorption
spectra using an ionizing background computed as a function of proper
time along the trajectory of a photon emitted by the quasar, rather
than at the proper time of the quasar. This effect is appropriate if
considering spectra at the end of the reionization era, when the
ionizing background can evolve significantly during the light travel
time across a quasar near-zone. The right hand panel of
Figure~\ref{fig2} illustrates the background photo-ionization rate as
a function of radius in our model (for a quasar observed at $z=6$).

\begin{figure}
\begin{center}
\vspace{-30mm}
\includegraphics[width=9cm]{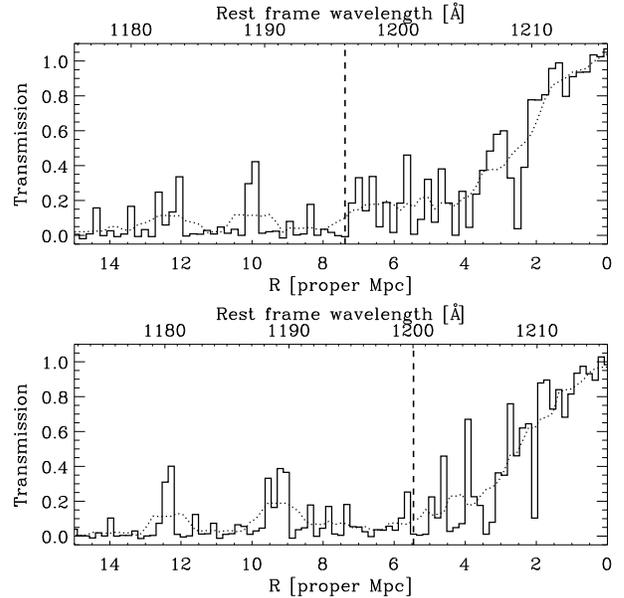}
\end{center}
\vspace{-7mm}
\caption{\label{fig3}Example synthetic near-zone spectra.  In both panels, the dotted curve shows the spectrum after being smoothed to a resolution of
20\AA~in the observed frame while the vertical dashed line marks the measured near-zone size.  This corresponds to the position where the smoothed flux first drops below a transmission of 10 per cent. {\em Top}: Quasar at $z=5.8918$ with $M_{1450}=-26.09$.  {\em Bottom}: Quasar at $z=6.247$ with $M_{1450}=-26.67$.}
\end{figure}

\begin{figure*}
\begin{center}
\includegraphics[width=13cm]{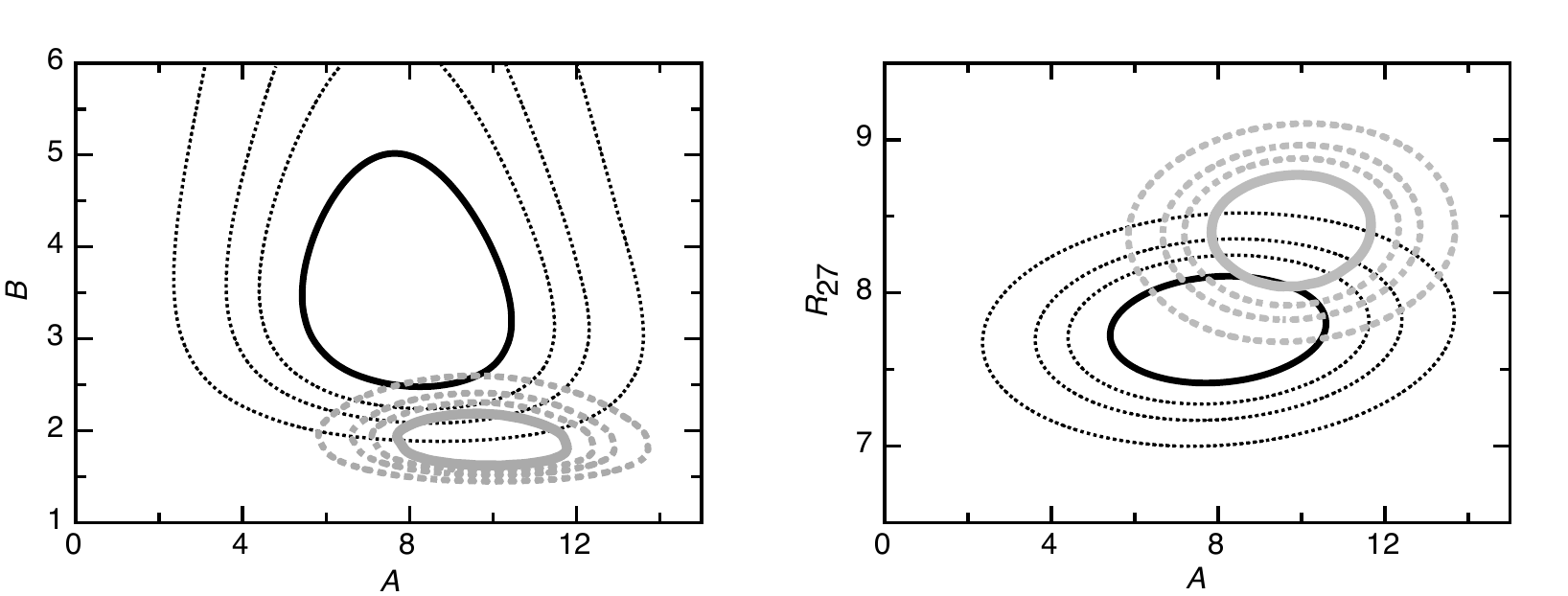}
\end{center}
\caption{\label{fig4} Projections of $\chi^2$ for  observed (black) and simulated (grey) near-zone sets assuming the fiducial evolving model.  We have introduced intrinsic scatter ($\sigma_{\rm obs}=1.86$ Mpc and $\sigma_{\rm sim}=1.27$ Mpc, respectively) so that the reduced
  $\chi^2$ is unity for the best fit model. The contours represent $\Delta\chi^2=1$, 2, 3 and
  5 relative to the corresponding minimum value of $\chi^2$. }
\end{figure*}

For each of the model near-zone sets constructed, the sizes of the
simulated near-zones were calculated for a sample of synthetic quasar
spectra with the same redshifts and quasar absolute magnitudes as the
\citet{carilli2010} compilation.  The synthetic spectra are designed
to match the spectral resolution ($R\sim 2500$) and signal to noise of
the observational data ($\rm S/N=20$ per $3.5\rm\,\AA$ pixel) as
closely as possible.  For the fiducial simulations an EUV spectral
index of $\alpha=1.5$ (where $dL/d\nu\propto\nu^{-\alpha}$) was
assumed to calculate the corresponding ionizing photon rate
($\dot{N}$) from the observed $M_{\rm 1450}$.  Additionally, a
temperature at mean density of $T=23600^{+5000}_{-6900}$~K has been
measured \citep{bolton2010} within the near-zone around SDSS
J0818+1722.  Our fiducial simulations are consistent with this
constraint, and we further assume this as the best fit temperature
when modelling near-zones throughout our analysis.  This level of
complexity is important as the near-zone size~\citep{fan2006} is
measured based on the first resolution element where the transmission
in the spectrum drops below 10 per cent, meaning that the Ly$\alpha$
forest must be simulated in detail in order for the modelling to be
considered reliable. Examples of some of the synthetic quasar spectra
used in our analysis are displayed in Figure~\ref{fig3}.

\section{Near-zone Relation Constraints}

\label{results}

As noted in the introduction, the appropriate value of $B$ describing
the power-law relation between near-zone size and luminosity is
uncertain. Therefore, rather than correct for luminosity by assuming a
fixed $B$, and fit equation~(\ref{nzrel}) to determine the evolution
of near-zone size with redshift as has been done previously, in our
paper we instead model the observed near-zone plane using free
parameters to describe both the evolution with redshift ($A$) and the
evolution with absolute magnitude ($B$),
\begin{equation}
\label{plane}
R_{\rm p} = R_{27} \times
  10^{-0.4(27+M_{\rm 1450})/ B} + A\times (z-6).
\end{equation}
In this section, we use equation~(\ref{plane}) to quantify constraints from the near-zone plane at $z\sim6$.

We find the parameter sets that describe the near-zone plane by
fitting equation~(\ref{plane}) to both the observed near-zone relation
and to our simulated data sets.  The black and grey contours in
Figure~\ref{fig4} show the resulting projections of $\chi^2$
onto the parameter spaces $( A, B)$ and $( A,R_{27})$ for the observed
and fiducial simulated near-zones, respectively. When calculating
$\chi^2$ we introduce intrinsic scatter into equation~(\ref{plane}) so
that the reduced $\chi^2$ of the best fit model is unity. We find a
smaller intrinsic scatter ($\sigma_{\rm sim}=1.27\pm0.04$ Mpc) in the
simulated data, indicating that sight-line variation is not
responsible for all of the observed scatter ($\sigma_{\rm
  obs}=1.86\pm0.06$).

While the simulations reproduce the size evolution with redshift, the
luminosity dependence is not well described, with $B\sim2$ for the
simulations (equation~\ref{nz}) but $ B\sim3$ for the observed
sample. There are two explanations for this disagreement between the
data and fiducial model which we now discuss in turn.

\section{Near-zone relation determined by HII regions}

\label{HIIregion}

Firstly, the observed value of $B=3$ is expected if the quasars were
surrounded by HII regions embedded in a neutral IGM~\citep{cen2000}.
Naively, this could be interpreted as evidence for a significantly
neutral IGM. To test this, we calculate the expected scatter in the
near-zone relation for comparison with observation.  In contrast to
the case for a proximity zone in an optically thin IGM, the size of an
HII region depends on the age of the quasar at the time of
observation. As a result, there is scatter in the observed size even
for fixed redshift and quasar luminosity.  We can estimate the scatter
due to the age of the quasars by noting that if there is equal
probability of observing a quasar at any age during its lifetime, then
the probability distribution for the HII region radius at fixed
luminosity $L$ and redshift $z$, given a quasar lifetime $t_{\rm max}$,
is
\begin{equation}
\left.\frac{dP}{dR_{\rm
      p}}\right|_{L,z}=\frac{3R_{\rm p}^2}{R_{\rm
      max}^3},
\end{equation}
where $R_{\rm max}$ is the size reached after $t_{\rm max}$. Given
this distribution, the mean is $\langle R_{\rm p}\rangle=3/4\,R_{\rm
  max}$, and the variance is $\sigma_R^2\equiv\langle R_{\rm
  p}^2\rangle - \langle R_{\rm p}\rangle^2 = \langle R_{\rm
  p}\rangle^2/15$.  Observationally, since we have $\langle R_{\rm
  p}\rangle\sim R_{27}\sim8$ Mpc, the scatter in HII region radius
owing to random quasar age is $\sigma_R\sim2$ Mpc. Thus the scatter in
observed quasar age accounts for all of the observed scatter
($\sigma_{\rm obs}=1.86\pm 0.06$) in the near-zone sizes. However in
the case of HII regions we would also expect additional scatter in the
size owing to other quantities like the quasar lifetime and the
spectral index (we return to this point below), as well as scatter due
to inhomogeneities in the density field and ionization structure along
the different quasar sight-lines. In the case of the latter, the sizes
of HII regions near the end of reionization~\citep{furlanetto2004} are
thought to vary over the range 1-2 proper Mpc, which would lead to a
component of scatter in addition to the $\sim2\,$Mpc expected from the
quasar age. Thus, the total scatter in an HII region generated
near-zone relation would be in excess of $\sigma_R\sim3$ Mpc.  We
therefore infer that although quasar HII regions would lead to a
near-zone relation with the appropriate value of $ B\sim3$, the
observed relation between near-zone size and magnitude would be too
tight in this case. This scatter based constraint would be alleviated if the near-zone
sizes corresponded to resonant absorption within an HII region (the
scatter would be smaller since the size is independent of lifetime in
this case). However, in this alternative case where resonant absorption
sets the near-zone size, our modelling suggests that $ B\sim2$
rather than $ B\sim3$ should be observed. As a result, HII regions cannot provide a viable
physical explanation for the observational data \citep[see also][]{maselli2009}.

\section{Constraining the EUV spectral index}

\label{EUV}

An alternative scenario to explain this trend is provided by an EUV
spectral index which is luminosity dependent, so that at brighter
absolute magnitudes the ionizing flux is smaller.  To quantify this
point, we first note that the observed quantities are the near-zone
size and the absolute UV magnitude. Given this UV magnitude, the
ionization properties of the surrounding medium are dependent on the
ionizing luminosity in photons per second ($\dot{N}$), and the
spectral index blueward of the Lyman break. The relation between
$\dot{N}$ and $\alpha$ is
\begin{equation}
\label{ndot}
\dot{N} = \int_{\nu_{\rm
      Ly}}^{x_{\rm max}\nu_{\rm Ly}}d\nu \frac{L_{\nu_{\rm
        Ly}}}{h\nu}\left(\frac{\nu}{\nu_{\rm
      Ly}}\right)^{-\alpha},
\end{equation}
where $L_{\nu_{\rm Ly}}$ is the luminosity at the Lyman limit
$\nu_{\rm Ly}$, $h$ is Planck's constant and $x_{\rm max}$ is the
ratio between the frequency where photons cease to contribute to
ionization and the Lyman limit (we take $x_{\rm max}=100$, though this
choice does not affect our results). To estimate the value of
$\dot{N}$ for the quasars in the \citet{carilli2010}
sample, the observed $M_{1450}$ has previously~\citep{bolton2010} been
combined with an assumed a spectral index of $\alpha=1.5$. Our
approach to investigate whether a variable spectral index is able to
explain the discrepancy in $ B$ is therefore to modify the value of
absolute magnitude that corresponds to the ionization rate at the
near-zone radius measured in the simulations. This
  allows us to investigate the near-zone relations that result from a
  range of spectral indicies without re-calculating the radiative
  transfer simulations.  Beginning with equation~(\ref{nz}) and
equation~(\ref{ndot}) and assuming a spectrum of the form
\begin{equation}
L_\nu (\nu) =
  L_{1450}\left(\frac{\nu_{1050}}{\nu_{1450}}\right)^{-0.5}
  \left(\frac{\nu_{\rm Ly}}{\nu_{1050}}\right)^{-\alpha}
  \left(\frac{\nu}{\nu_{\rm Ly}}\right)^{-\alpha},
\end{equation}
we
adjust the absolute magnitude corresponding to the model quasars (at
fixed near-zone radius) for a variation in spectral index
using
\begin{eqnarray}
\label{Mdif}
\nonumber
M_{\rm 1450} &=&
  M_{\rm 1450,1.5} -
  2.5\log_{10}\left(\left(\frac{\nu_{912}}{\nu_{1050}}\right)^{\alpha-1.5}
  \frac{\alpha+3}{4.5}\right.\\
&&\hspace{-18mm}\times\left.\frac{1-x_{\rm
      max}^{-4.5}}{1-x_{\rm max}^{-(\alpha+3)}}
  \left(1.71-0.86\alpha+0.33\alpha^2-0.05\alpha^3\right)^{-0.7}
  \right),
\end{eqnarray}
where $M_{\rm b,1.5}$ is the observed
absolute magnitude used in the fiducial model. Here we note that
simulations performed with $\alpha=0.5$, $\alpha=1.0$, $\alpha=1.5$, $\alpha=2.0$, and
$\alpha=2.5$ yield near-zone temperatures of $T\sim2.81\times10^4$K,
$T\sim2.34\times10^4$K , $T\sim2.06\times10^4$K ,
$T\sim1.89\times10^4$K and $T\sim1.76\times10^4$K respectively. These
values are fit (to within 1\%) by
\begin{equation}
T =
  35400-17700\alpha+6757.1\alpha^2-1000\alpha^3.
\end{equation}
This dependence
contributes the last product in equation~(\ref{Mdif}).

\begin{figure*}
\begin{center}
\includegraphics[width=13cm]{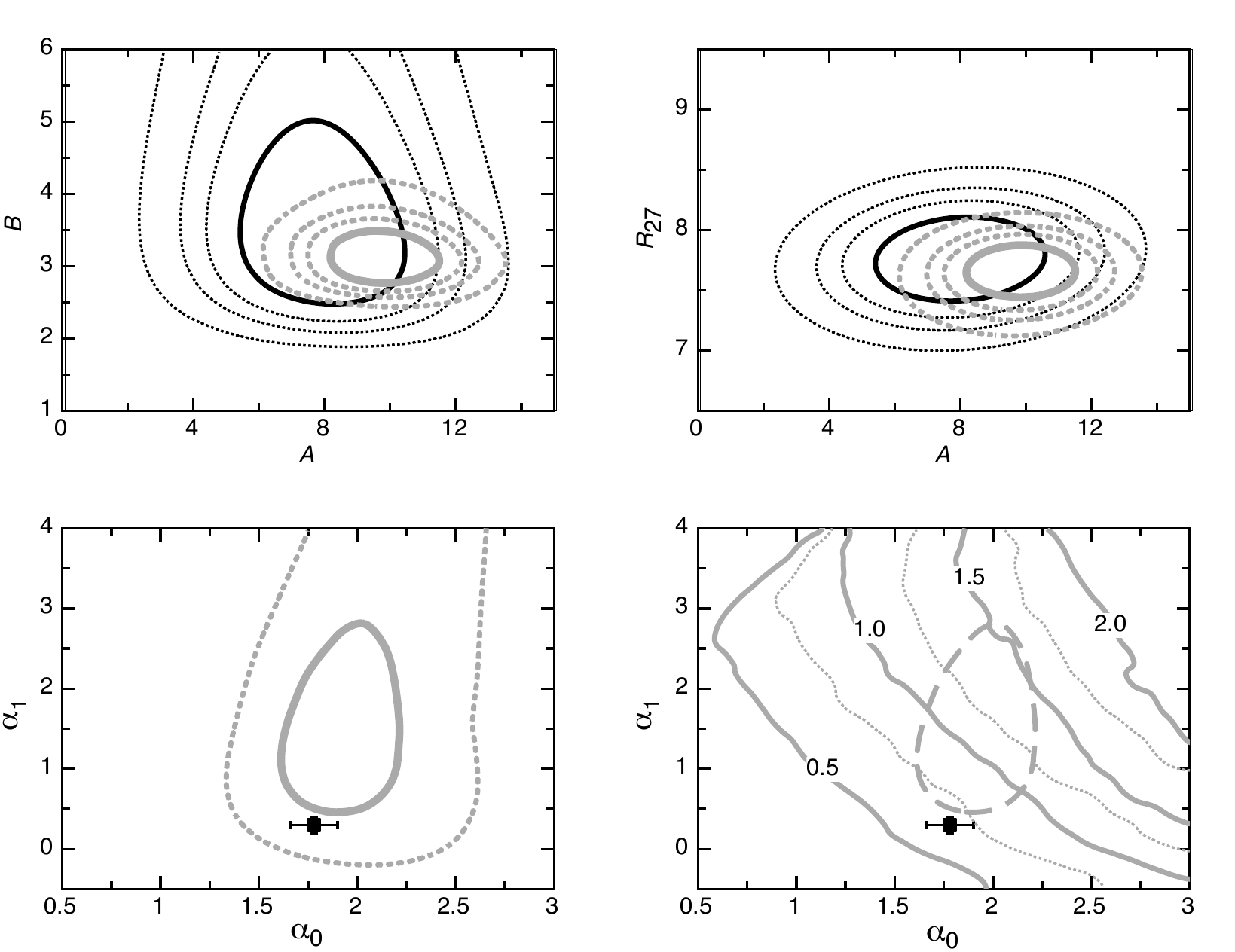}
\end{center}
\caption{\label{fig5}{\em Upper panels:} Projections of $\chi^2$ for  observed (black) and simulated (grey) near-zone sets for the best fit  simulations (in $\alpha_0$ and $\alpha_1$) assuming the fiducial evolving
  ionizing background.  We have introduced intrinsic scatter so that the reduced
  $\chi^2$ is unity for the best fit model. The contours represent $\Delta\chi^2=1$, 2, 3 and
  5 relative to the corresponding minimum value of $\chi^2$. {\em Lower left panel}: Contours of likelihood for the values of $\alpha_0$ and $\alpha_1$ (at 60\% and 10\% of
  the peak likelihood). {\em Lower right panel}: Contours of scatter in spectral index $\alpha$. The point with error-bars shows the corresponding parameters measured~\citep{telfer2002} at low redshift. }
\end{figure*}

To quantify the
possible dependence of spectral index on luminosity we parametrize
the value of $\alpha$ as a function of $M_{\rm 1450}$
using
\begin{equation}
\alpha = \alpha_0 - \alpha_1(M^\prime_{\rm
    1450}+27),
\end{equation} where $\alpha_0$ represents the mean
spectral slope of quasars with $M_{1450}=-27$ and $\alpha_1=
\frac{d\alpha}{dM_{\rm 1450}}$ describes the variation of spectral
index with magnitude. Note that since $\alpha$ is a function of
$M_{1450}$, we evaluate it not at the observed $M_{1450}$, but instead
use an absolute magnitude $M^\prime_{1450}$ that is adjusted to
account for the average index $\alpha_0$ at $M_{1450}=-27$,
\begin{eqnarray}
\nonumber
M^\prime_{\rm 1450} &=&
  M_{\rm 1450,1.5} -
  2.5\log_{10}\left(\left(\frac{\nu_{912}}{\nu_{1050}}\right)^{\alpha_0-1.5}
  \frac{\alpha_0+3}{4.5}\right.\\
&&\hspace{-18mm}\times\left.\frac{1-x_{\rm
      max}^{-4.5}}{1-x_{\rm max}^{-(\alpha_0+3)}}
  \left(1.71-0.86\alpha_0+0.33\alpha_0^2-0.05\alpha_0^3\right)^{-0.7}
  \right).
\end{eqnarray}
We use these equations to constrain the
likelihood of parameter combinations $(\alpha_0,\alpha_1)$ through convolution
of the resulting joint probability distribution for the parameters $
A$, $ B$ and $R_{27}$ that describe simulated near-zone relations,
with the corresponding relation for the
observations
\begin{equation}
\nonumber
\mathcal{L}_{\alpha_0,\alpha_1}
  \propto \int dA \, dB \,dR_{27} \frac{d^3P_{\rm
      obs}}{dA\,dB\,dR_{27}}\left.\frac{d^3P_{\rm
      sim}}{dA\,dB\,dR_{27}}\right|_{\alpha_0,\alpha_1} \mathcal{L}_{\sigma_{\rm
      sim}},
\end{equation}
where
\begin{equation}
\nonumber
\frac{d^3P_{\rm obs}}{dA\,dB\,dR_{27}}\propto
  e^{-\frac{\chi_{\rm obs}^2(A,B,R_{27})}{2}},
\end{equation}
and
\begin{equation}
\frac{d^3P_{\rm sim}}{dA\,dB\,dR_{27}}\propto
  e^{-\frac{\chi_{\rm sim}^2(A,B,R_{27})}{2}}.
\end{equation}
Here
we have assumed flat prior probability distributions for each of the
parameters $A$, $B$ and $R_{27}$. As part of this procedure we
calculate the intrinsic scatter for each set of parameters
$(\alpha_0,\alpha_1)$, and impose the condition that this scatter be smaller
than the observed scatter by weighting the calculated likelihood
with
\begin{eqnarray}
\nonumber
 \mathcal{L}_{\sigma_{\rm
      sim}}&\propto&e^\frac{-(\sigma_{\rm obs}-\sigma_{\rm sim})^2}{2
    (0.06^2)},\hspace{3mm}\mbox{where}\hspace{3mm} \sigma_{\rm
    obs}<\sigma_{\rm sim}\\
  &\propto&1\hspace{3mm}\mbox{otherwise.}
 \end{eqnarray}
Here the
value of 0.06 dex represents the uncertainty in $\sigma_{\rm obs}$.

\begin{figure*}
\begin{center}
\includegraphics[width=13cm]{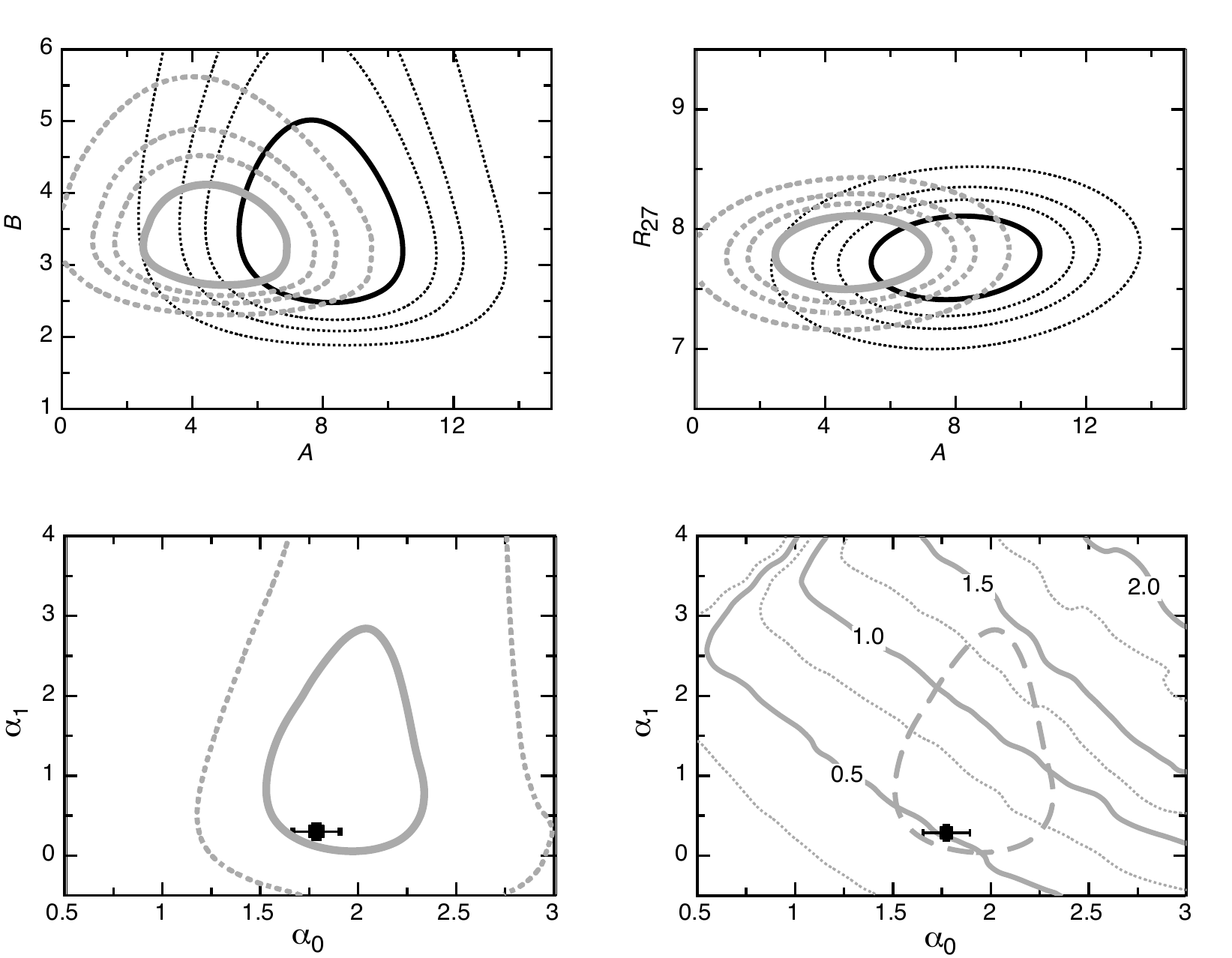}
\end{center}
\caption{\label{fig6}{\em Upper panels:} Projections of $\chi^2$
  for observed (black) and simulated (grey) near-zone sets for the
  best fit simulations (in $\alpha_0$ and $\alpha_1$) assuming the
   non-evolving ionizing background.  We have introduced
  intrinsic scatter so that the reduced $\chi^2$ is unity for the best
  fit model. The contours represent $\Delta\chi^2=1$, 2, 3 and 5
  relative to the corresponding minimum value for $\chi^2$. {\em Lower
    left panel}: Contours of likelihood for the values of $\alpha_0$
  and $\alpha_1$ (at 60\% and 10\% of the peak likelihood). {\em Lower
    right panel}: Contours of scatter in spectral index $\alpha$. The
  point with error-bars shows the corresponding parameters
  measured~\citep{telfer2002} at low redshift. }
\end{figure*}

Contours of likelihood for the values of $\alpha_0$ and $\alpha_1$ based on our fiducial model are plotted
in the lower left panel of Figure~\ref{fig5}. We find that the preferred value
of $\alpha_0$ is smaller than the assumed $\alpha=1.5$. In addition, since the
fiducial simulations under-predict $ B$ (Figure~\ref{fig4}), the value
of $\alpha_1$ is found to be greater than zero with high confidence, so
that more luminous quasars have softer spectra.  This is consistent with the results of \citet[][]{strateva2005}, who found that brighter quasars are relatively less luminous in the X-ray, and is also expected from theoretical modeling of the quasar spectrum~\citep[][]{wandel88}.

The corresponding
contours (grey curves) of $\Delta\chi^2$ in the parameter spaces $(
A, B)$ and $( A,R_{27})$ for the simulated near-zone relations computed using the most likely values of  $\alpha_0$ and $\alpha_1$
are shown in the upper panels of Figure~\ref{fig5}, along with
contours for the observed near-zone sample (black curves). The best
fit simulations reproduce the observed values of average size $R_{27}$
and index $ B$. The parameter $ A$, which is sensitive to the
evolution of the ionizing background, is also consistent with
observations~\citep{fan2006,bolton2007b}.  The near-zone sizes for the
best fit model are plotted in
Figure~\ref{fig1} (triangles). 

\subsection{Intrinsic scatter in the spectral index}

As mentioned previously in section 5, the intrinsic scatter in the
simulated near-zones ($\sigma_{\rm sim}=1.25$ for the best fit model)
is smaller than the scatter $\sigma_{\rm obs}=1.86$ in the observed
near-zone relation. Since the numerical simulations already include
line-of-sight density variations in the IGM, the missing scatter
\begin{equation}
\Delta R_{\rm p} = \sqrt{\sigma_{\rm obs}^2-\sigma_{\rm sim}^2}
\end{equation}
could be provided by scatter in the spectral index among individual quasars, which is known to be significant at lower redshift \citep[][]{telfer2002}. Using the relation for the missing scatter in $R_{\rm p}$ we obtain the magnitude of scatter in the spectral index that is necessary to simulate the observed near-zone scatter 
\begin{equation}
\Delta \alpha = \frac{d\alpha}{dM}\frac{dM}{dR}\Delta R= \alpha\frac{ B}{R_{27}}\sqrt{\sigma_{\rm obs}^2-\sigma_{\rm sim}^2}.
\end{equation}
Contours of this scatter are shown in the right hand panel of Figure~\ref{fig5}. Values for scatter in spectral index ranging between $\Delta \alpha\sim0.75$ and $\Delta \alpha\sim1.25$ correspond to the range of parameters $\alpha_0$ and $\alpha_1$  which lead to models for the near-zone relation that are in good agreement with the data.

\subsection{Constant ionizing background models}

\begin{figure*}
\begin{center}
\includegraphics[width=17cm]{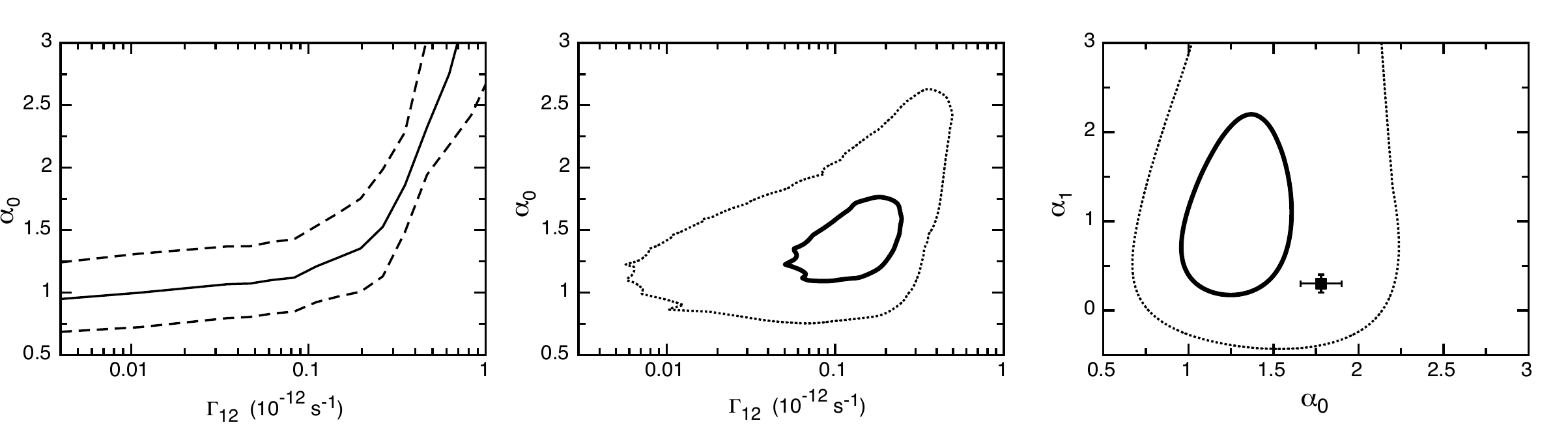}
\end{center}
\caption{\label{fig7}{\em Left Panel}: The best-fit value
  for $\alpha_0$ as a function of $\Gamma_{12}$ (solid curve),
  together with the 68\% range (dashed curves). {\em Central Panel}: The
  joint probability distribution for $\Gamma_{12}$ and $\alpha_0$. The
  prior probability for $\Gamma_{12}$ is assumed to be flat in the
  logarithm ($dP_{\rm prior}/d\Gamma_{12}\propto 1/\Gamma_{12}$).
  {\em Right Panel}: The joint probability distribution for $\alpha_0$ and
  $\alpha_1$. In the central and right panels the contours represent levels
  that are at 60\% and 10\% of the peak likelihood.  The final
  estimate for the EUV spectral index is $\alpha_0=1.3^{+0.4}_{-0.3}$. }
\end{figure*}

Our simulations of quasar near-zones assume a semi-analytic model for
the ionizing background \citep[see ][]{wyithe2008}. It is therefore
important to ask whether the results in this paper pertaining to the
dependence of near-zone size on luminosity (i.e. the constraints on
$B$), and the resulting conclusions regarding the quasar spectral
index $\alpha$ are sensitive to this model. To address this issue we
have repeated our modelling of quasar near-zones assuming an ionizing
background that is independent of redshift. We first take the $z=6$
value from our fiducial model; Figure~\ref{fig6} shows contours
describing the resulting constraints (grey curves). The results are
very similar to the fiducial case, with the exception of $ A$, which
shows that the redshift evolution of near-zone size in the constant
background model is smaller than observed.  This is consistent with
the previous inference that the trend of near-zone size with redshift
is being driven by the rising intensity of the ionizing
background~\citep{wyithe2008} at $z\sim 6$.  We have also tested
sensitivity to the ionizing background amplitude by repeating our
analysis for 16 different values ranging between 1/100 and $\sqrt{10}$
times the fiducial model. In the left panel of Figure~\ref{fig7} we
show the resulting range for $\alpha_0$ as a function of the
background photoionization rate $\Gamma_{12}=\Gamma_{\rm
  HI}/10^{-12}\rm\,s^{-1}$. This figure illustrates that results for
$\alpha$ depend on the assumed value of $\Gamma_{12}$. As a result, we next
constrain $\Gamma_{12}$ following the work of \citet{bolton2007b}.

\subsubsection{The Ionizing Background at $z=6$}
\label{IBG}

\begin{table}
\begin{center}
\begin{tabular}{ccc}
\hline\hline
$z$
      & $\mathcal{T}$ & $\sigma_\mathcal{T}$ \\
\hline 
4.92 &
      0.1276 & 0.0011 \\
4.95 & 0.1139 & 0.0022 \\ 
4.98 & 0.1002 &
      0.0050 \\
5.01 & 0.1268 & 0.0117 \\
5.04 & 0.1567 & 0.0073
      \\
5.06 & 0.1765 & 0.0190 \\
5.06 & 0.1285 & 0.0042 \\
5.06 &
      0.1509 & 0.0224 \\
5.07 & 0.0751 & 0.0011 \\
5.08 & 0.0530 &
      0.0025 \\
5.10 & 0.0898 & 0.0020 \\
5.11 & 0.1650 & 0.0030
      \\
5.13 & 0.1293 & 0.0064 \\
5.13 & 0.1243 & 0.0050 \\\\
5.90
      & 0.0108 & 0.0033 \\
5.93 & 0.0125 & 0.0022 \\
5.95 & 0.0038 &
      0.0005 \\
5.95 & 0.0060 & 0.0010 \\
6.08 & -0.0071 & 0.0020
      \\
6.10 & 0.0012 & 0.0010 \\
6.10 & 0.0051 & 0.0005 \\
6.25 &
      0.0015 & 0.0005
      \\
\hline
\end{tabular}
\end{center}
\caption{\label{Table1}Table of transmissions $\mathcal{T}$ and uncertainty
$\sigma_\mathcal{T}$ in $\Delta z=0.1$ regions from \citet[][]{fan2006}.}
\end{table}

\citet{fan2006} present eight
values of transmission $\mathcal{T}$ (with uncertainty
$\sigma_\mathcal{T}$) measured in redshift intervals of $\Delta z=0.1$
centred on redshifts in the range $5.9\leq z\leq6.25$, and 14 values
centred in the redshift range $4.9\leq z\leq5.15$. These values are
listed in Table~\ref{Table1}. From these values we calculate the mean
value of transmission $\langle \mathcal{T}\rangle=0.004\pm0.002$ and
$\langle \mathcal{T}\rangle=0.12\pm0.01$ respectively.  To estimate
this uncertainty we have calculated the standard error on the mean
transmission via a bootstrap analysis which includes the uncertainties
on individual parameters. The values of $\langle
\mathcal{T}\rangle=0.004$ and $\langle \mathcal{T}\rangle=0.12$ are
consistent with the values quoted previously~\citep{bolton2007} at
$z=6$ and $z=5$ respectively, although note that this study uses
slightly wider redshift bins and does not use the independent data
of \citet{songaila2004}. However our estimates of the uncertainty are
smaller, particularly at $z=6$ where the previously quoted error
\citep{bolton2007} is a factor of $\sim2$ larger. This difference can
be attributed to the fact that rather than estimate the standard error
on the mean transmission, \citet{bolton2007} used an
inter-quartile range as an estimate for the error on the transmission
values along various lines of sight.

To estimate the corresponding background photo-ionization rate we
next combine previously published~\citep{bolton2007} data with a
Monte-Carlo method. We first determine the resulting distribution of
effective optical depths from our transmission estimates, finding
values of $\tau_{\rm eff}=5.5_{-0.4}^{+0.6}$ and $\tau_{\rm
  eff}=2.1_{-0.07}^{+0.08}$ at $z=6$ and $z=5$ respectively.  Here and
below we have quoted the median and the 68\% range. We note that the
value of $\tau_{\rm eff}$ is not defined in the small number of cases
where the Monte-Carlo trial value of $\mathcal{T}$ is negative. In
such cases we set $\mathcal{T}=10^{-6}$ and note that our results are
insensitive to the exact value provided it is well below the
observational limits in the Gunn-Peterson troughs.
Given the
distribution of effective optical depths, we determine the
corresponding probability distribution for the normalized
photoionization rate $\Gamma_{12}=\Gamma_{\rm
  HI}/10^{-12}\rm\,s^{-1}$. To estimate the uncertainties we have
utilized scaling relations derived from hydrodynamical simulations \citep[see Table 4 of ][]{bolton2007} for $\Gamma_{12}$, as functions of
$\tau_{\rm eff}$, the IGM temperature [which we take to be
$\log_{10}{(T/10^4\mbox{K})}=0\pm0.3$], and the slope of the temperature density
relation. We also utilize scaling relations as a function of the cosmological parameters. 

At $z=5$ we find
$\Gamma_{12}=0.47_{-0.2}^{+0.3}$, while at $z=6$ we obtain
$\Gamma_{12}=0.18_{-0.09}^{+0.18}$. We note that the smaller
uncertainty in $\tau_{\rm eff}$ relative to the previous estimate
~\citep{bolton2007} allows us to quote a probability density for the
measured ionizing background (with a non-zero value preferred at
2$\sigma$) rather than an upper limit. The revised estimates of the
ionizing background at $z=5$ and $z=6$ are presented in Figure~\ref{fig2}.

\subsubsection{Joint Probability for $\Gamma_{12}$ and $\alpha_0$}

We
use the constraints for $\Gamma_{12}$, together with the likelihoods for $\alpha_0$ and $\alpha_1$
as a function of $\Gamma_{12}$, to generate a joint probability
distribution for $\Gamma_{12}$ and $\alpha_0$ (central panel of
Figure~\ref{fig7}). Similarly, we obtain the joint probability for $\alpha_0$
and $\alpha_1$ (right hand panel of Figure~\ref{fig7}). This procedure yields an
estimate of $\alpha_0=1.3^{+0.4}_{-0.3}$. This constraint represents the primary measurement of our paper.

\subsection{Comparison with spectral index at low redshift}

\begin{figure*}
\begin{center}
\includegraphics[width=17cm]{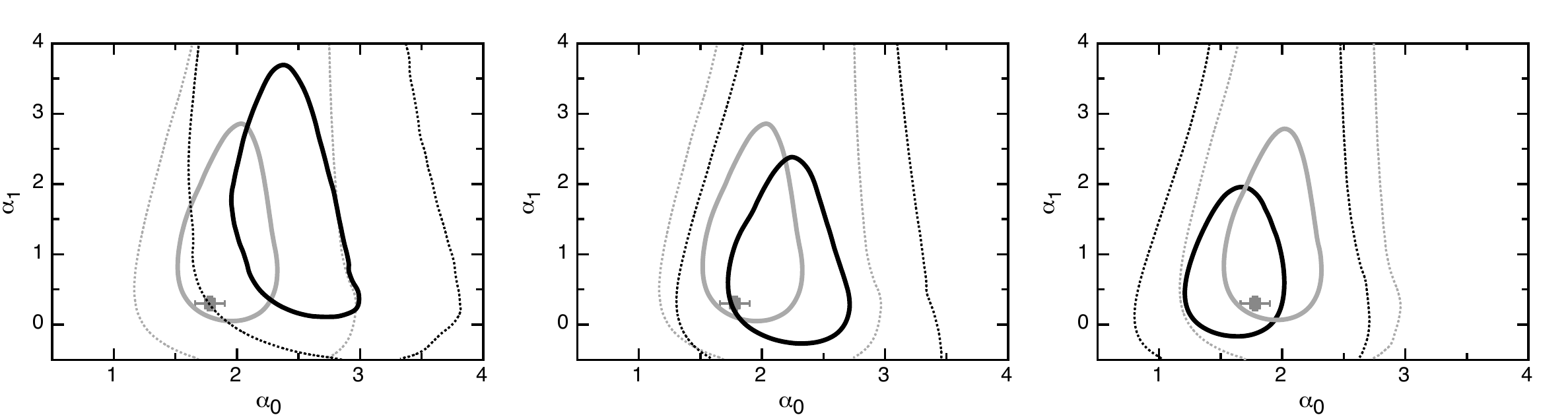}
\end{center}
\caption{\label{fig8}Figure showing the systematic uncertainties on
  the joint probability distribution for $\alpha_0$ and
  $\alpha_1$. {\em Left panel:} The dark contours show constraints
  that have been computed based on a model with $\alpha=0.5$. {\em
    Central panel:} The dark contours show the constraints in a case
  where the continuum has been deliberately placed at a level that is
  5\% too low. {\em Right panel:} The dark contours show the
  constraints in a case where where self-shielded absorbers have been
  added to regions with density contrast $\Delta>16$. In each case the
  model assumes an unevolving ionizing background, for which the light
  contours in each panel show the constraints. The contours represent
  levels that are at 60\% and 10\% of the peak likelihood
  respectively.}
\end{figure*}

\citet[][]{telfer2002} measured the UV spectral index for quasars at
$z\sim1-2$. By analysing the composite spectrum they found a mean
value of $\langle \alpha\rangle=1.76\pm0.12$. \citet[][]{telfer2002}
also measured $\alpha$ for a sub-sample of individual quasars,
allowing them to estimate the variation $\alpha_1=0.3\pm0.1$ with
luminosity\footnote{Note that the evolution of $\alpha$ with
  luminosity is dominated by radio loud quasars, which formed half the
  total sample of quasars for which individual measurements of
  $\alpha$ were made by \citet[][]{telfer2002}.}, as well as the
intrinsic scatter $\Delta \alpha=0.76$. These estimates are plotted
alongside our model constraints in
Figure~\ref{fig5}-\ref{fig8}. We find that the parameter
combination $\alpha_0=1.76\pm0.12$ and $\alpha_1=0.3\pm0.1$ is
consistent with our inference for these parameters at
$z\sim6$. Moreover, we find that our inferred scatter in $\alpha$ at
$z\sim6$ agrees well with the $\Delta \alpha=0.76$ found by
\citet[][]{telfer2002}. Thus, assuming our fiducial model of the
ionizing background, there is no evidence for evolution in the EUV
slope of quasars with redshift out to $z\sim6$.

\section{Uncertainties}
\label{other}

In addition to those already discussed, we have checked several other
sources of uncertainty. Specifically, we have recalculated our constraints including the assumption of a fiducial model
with $\alpha=0.5$ rather than $\alpha=1.5$, a model where the quasar
continuum has been placed systematically high by 5\% on the data, and
a model that includes an approximate treatment of Lyman limit
systems. Each of these effects is found to have a relatively small
effect on the derived parameters for the spectral index relative to
the quoted uncertainty, as detailed below.

\subsection{Near-zone temperature}
There is an uncertainty in the absolute temperature of the IGM within
the quasar near-zone, as well as scatter in the temperature from one
near-zone to another. These uncertainties will impact on our
constraints for the EUV spectral index. Firstly, any scatter in
temperature among different near-zones will contribute to scatter in the
observed near-zone sizes. However this scatter will be degenerate with
scatter in the spectral index ($\Delta \alpha$). Thus the potential presence of scatter
in temperature does not introduce uncertainty into our estimated mean
value of $\alpha_0$ or $\alpha_1$.  However, it does imply that our quoted
$\Delta \alpha$ should be considered an upper limit.

On the other hand, uncertainty in the absolute value of the
temperature within the near-zones could introduce a more serious
error. The temperature of the IGM in the quasar near-zone has two
independent contributions which add linearly, and hence the near-zone
temperature has two components of uncertainty. Firstly, the mean IGM
has an uncertainty in temperature resulting from heating during
hydrogen reionization.  This component of uncertainty in the temperature
due to hydrogen reionization is already included~\citep{bolton2007} in
the calculation of the ionizing background.

However in addition the reionization of HeII by the quasar may
heat the IGM in the near-zone to well above this mean IGM
temperature~\citep{boltonoh2009}.  In our modelling we have assumed a
temperature~\citep{bolton2010} that is consistent with observations of
a single $z\sim6$ quasar (SDSS J0818+1722), and that the quasar is
responsible for reionization of HeII within the near-zone.  This
latter assumption means the temperature increase from HeII
reionization by the quasar is directly related to the spectral
hardness of the quasar, and hence to $\alpha$, which is the quantity we are
trying to measure. As a result, the assumption of an incorrect
temperature could bias the inferred value of $\alpha$ in our analysis. A
hardening of the spectral index leads both to higher temperatures
within the near-zone and to an enhanced photo-ionization rate.  By
assuming a fixed temperature for all modelling our analysis would
include only the latter enhancement, whereas both ionization and
temperature influence the optical depth. We have therefore accounted
for this effect via the scaling in equation~(\ref{Mdif}) based on a
calibration of near-zone temperature with spectral index estimated
from additional radiative transfer simulations.  Furthermore, to check
the importance of the temperature effects, and at the same time check
that our assumed fiducial value of $\alpha$ does not influence the results,
we have repeated the constraints on $\alpha_0$ and $\alpha_1$ based on the
fiducial unevolving ionizing background using a simulated quasar with
a value of $\alpha=0.5$ (and correspondingly hotter temperature). As shown in the left panel of Figure~\ref{fig8}, we find
that the assumed fiducial $\alpha$ (and corresponding fiducial temperature)
does not strongly influence the constraints.

\subsection{Continuum placement}

In addition to errors in the redshifts of the quasars,
uncertainty in continuum placement may also lead to an error in the
near-zone radius $R_{\rm p}$. This error corresponds to an
uncertainty \citep{fan2006} of $\Delta \tau\sim0.05$,
or equivalently to a relative uncertainty in the transmission of
$\Delta\mathcal{T}/\mathcal{T}\sim5\%$. At the near-zone radius
$\mathcal{T}=10\%$, so that the uncertainty in transmission is
$\Delta\mathcal{T}\sim0.005$. Inspection of Figure~3 in \citet{carilli2010} shows that when averaged over the near-zone
sample, the gradient of transmission with radius is
$d\mathcal{T}/dR_{\rm p}\sim0.05$ Mpc$^{-1}$. Hence the uncertainty in
the near-zone radius that is introduced by the uncertainty in continuum
placement is $\Delta R_{\rm p}\sim dR_{\rm
  p}/d\mathcal{T}\Delta\mathcal{T}\sim0.1$Mpc. Thus, continuum
placement is sub-dominant in terms of contribution to the
uncertainty in near-zone size. We note that the uncertainty could be
larger for individual quasars. However this would introduce additional
scatter into the near-zone relation which would be degenerate with the
other sources of scatter. To check this conclusion we have again
repeated the constraints on $\alpha_0$ and $\alpha_1$ based on an
unevolving ionizing background using simulated quasars for which the
continuum has been systematically placed 5 per cent too low. As shown in the central panel of Figure~\ref{fig8}, we find that
the continuum placement has a systematic effect on near-zone size and
hence $\alpha_0$. However the shift is much smaller than the statistical
uncertainty. We thus conclude that our results are robust to a
systematic uncertainty in the continuum at this level.

\subsection{Lyman limit systems}

Finally, a potential shortcoming of our modelling is the assumption
that the IGM is optically thin, which results in the absence of
absorbers that are self-shielded with respect to the ionizing
background. To address this we have tested the sensitivity of
near-zone size by adding a population of self-shielded Ly-limit
systems. We achieve this by setting all gas with density contrast
$\Delta>16$~\citep[corresponding to the density contrast of a clump that
will self shield if immersed in our fiducial ionizing
background,][]{FurlanettoOh05} to be neutral prior to the quasar
turning on~\citep{bolton2007}. We find that their inclusion does not
alter the predicted distribution of near-zone sizes in most cases, and
that as shown in Figure~\ref{fig8}, the presence of self-shielded
systems should not significantly modify our measurement of $\alpha_0$
relative to other uncertainties. 

A related question concerns an
alternative explanation for the observed $B=3$ due to an enhanced
number density of absorbing systems in the biased environments
surrounding luminous quasars. Such an enhancement could preferentially
retard the observed near-zone sizes, leading to an increase in
near-zone size with luminosity that is slower than expected in the
optically thin regime. However we have previously~\citep{bolton2007}
investigated the dependence of near-zone size on quasar host-halo
mass, finding a negligible effect. This is because the absorbers that
set the near-zone size are located between 30 and 100 co-moving Mpc
from the quasar. Thus, the environmental dependence of absorber
abundance is unlikely to explain the observed $B=3$.

\section{conclusion}
\label{conclusion}

The discovery of a strong correlation of the sizes of near-zones
around high redshift quasars and their redshift in the range $5.7\la
z\la6.4$ \citep[][]{fan2006,carilli2010} has prompted a series of
studies aimed at understanding its implications for the tail end of
reionization
\citep[][]{fan2006,bolton2007,wyithe2008,bolton2010,carilli2010}. In
this paper we have noted that the amplitude of the near-zone radius
relation is sensitive to the assumed EUV spectral index of the
quasar. We therefore performed a large suite of numerical simulations
of the near-zone spectra using radiative transfer through a
hydro-dynamical model of the IGM in the quasar environment, and find
that the observed near-zone relation can be used to constrain the UV
spectral index of $z\sim6$ quasars. We find that the typical value of
spectral index for luminous quasars at $z\sim6$ is
$\alpha=1.3^{+0.4}_{-0.3}$ (for a specific luminosity of the form
$L_\nu\propto\nu^{-\alpha}$), where the uncertainty includes both the
statistical uncertainty in the near-zone relation as well as
uncertainty in the ionizing background. Based on comparison of the
scatter in our simulations and the scatter in the observed near-zone
relation we infer that the scatter in the spectral index is $\Delta
\alpha\sim0.75-1.25$.

\citet[][]{carilli2010} noted that the data are consistent with
near-zone sizes that scale with quasar luminosity as $R_{\rm p}\propto
L^{1/ B}$, where $B=3$, and that the scatter in the near-zone
radius--redshift relation is reduced if this scaling is
applied. However as pointed out by \citet[][]{bolton2007}, the value
of the power-law index $ B$ describing the scaling of near-zone radius
with luminosity that is physically appropriate depends on whether
near-zone sizes are set by HII region boundaries ($ B=3$) or resonant
absorption in an ionized IGM ($ B=2$). Motivated by the possibility of
using this second parameter in the near-zone relation to gain a better
understanding of its meaning, in this paper we jointly fit the
available observations for variation of near-zone size with both
redshift and luminosity. From this analysis we find that the
observations have an index of $ B\sim3$, with an uncertainty that
excludes a value of $ B=2$ with high confidence. In contrast to the
observations, we find that our fiducial numerical simulations predict a
value of $B=2$.

We have discussed two possibilities for this difference. Firstly our
simulations are conducted in a highly ionized IGM. It is possible that
although these simulations are able to describe the size and evolution
of the near-zones \citep[][]{bolton2010}, they do not represent
physical conditions of the IGM at $z\sim6$, which could instead
contain hydrogen that is significantly neutral as previously argued in
\citet[][]{wyithe2004} and \citet[][]{wyithe2005}. Although this would
lead to the observed $B=3$, we do not believe that this is the case
because the observed relation is too tight to be explained by HII
regions, the sizes of which contain a range of additional sources of
scatter. Instead, we argue that the observed value of $ B\sim3$ is
evidence for a UV spectral index that varies with absolute magnitude
in the high redshift quasar sample, becoming softer at higher
luminosity. This finding is in agreement with theoretical modeling of quasar spectra~\citep[][]{wandel88}.

The results of this paper provide the first constraints on the
properties of the EUV spectral index of the highest redshift quasars,
which are not observable owing to absorption blueward of the Lyman
limit. The values we find are in agreement with direct observations at
low redshift, and therefore indicate that there is no evidence for
evolution in the EUV properties of quasars over most of cosmic
time. The new constraints will aid all studies of reionization
involving high redshift quasars, which have previously relied on an
assumed value of the mean EUV spectral index measured at lower
redshift.

\section*{Acknowledgments}
This work was supported in part by the Australian Research Council. We thank Chris Carilli and Martin Haehnelt for helpful comments. 

\newcommand{\noopsort}[1]{}

\bibliographystyle{mn2e}
\bibliography{text}

\begin{thebibliography}{}

\bibitem[\protect\citeauthoryear{{Bolton}, {Becker}, {Wyithe}, {Haehnelt} \&
  {Sargent}}{{Bolton} et~al.}{2010}]{bolton2010}
{Bolton} J.~S.,  {Becker} G.~D.,  {Wyithe} J.~S.~B.,  {Haehnelt} M.~G.,
  {Sargent} W.~L.~W.,  2010, ArXiv e-prints, 10013415

\bibitem[\protect\citeauthoryear{{Bolton} \& {Haehnelt}}{{Bolton} \&
  {Haehnelt}}{2007a}]{bolton2007}
{Bolton} J.~S.,  {Haehnelt} M.~G.,  2007a, \mnras, 374, 493

\bibitem[\protect\citeauthoryear{{Bolton} \& {Haehnelt}}{{Bolton} \&
  {Haehnelt}}{2007b}]{bolton2007b}
{Bolton} J.~S.,  {Haehnelt} M.~G.,  2007b, \mnras, 382, 325

\bibitem[\protect\citeauthoryear{{Bolton}, {Oh} \& {Furlanetto}}{{Bolton}
  et~al.}{2009}]{boltonoh2009}
{Bolton} J.~S.,  {Oh} S.~P.,    {Furlanetto} S.~R.,  2009, \mnras, 395, 736

\bibitem[\protect\citeauthoryear{{Brandt}, {Schneider}, {Fan}, {Strauss},
  {Gunn}, {Richards}, {Anderson}, {Vanden Berk}, {Bahcall}, {Brinkmann},
  {Brunner}, {Chen}, {Hennessy}, {Lamb}, {Voges} \& {York}}{{Brandt}
  et~al.}{2002}]{brandt2002}
{Brandt} W.~N.,  {Schneider} D.~P.,  {Fan} X.,  {Strauss} M.~A.,  {Gunn} J.~E.,
   {Richards} G.~T.,  {Anderson} S.~F.,  {Vanden Berk} D.~E.,  {Bahcall} N.~A.,
   {Brinkmann} J.,  {Brunner} R.,  {Chen} B.,  {Hennessy} G.~S.,  {Lamb} D.~Q.,
   {Voges} W.,    {York} D.~G.,  2002, \apjl, 569, L5

\bibitem[\protect\citeauthoryear{{Carilli}, {Wang}, {Fan}, {Walter}, {Kurk},
  {Riechers}, {Wagg}, {Hennawi}, {Jiang}, {Menten}, {Bertoldi}, {Strauss} \&
  {Cox}}{{Carilli} et~al.}{2010}]{carilli2010}
{Carilli} C.~L.,  {Wang} R.,  {Fan} X.,  {Walter} F.,  {Kurk} J.,  {Riechers}
  D.,  {Wagg} J.,  {Hennawi} J.,  {Jiang} L.,  {Menten} K.~M.,  {Bertoldi} F.,
  {Strauss} M.~A.,    {Cox} P.,  2010, ArXiv e-prints, 10030016

\bibitem[\protect\citeauthoryear{{Cen} \& {Haiman}}{{Cen} \&
  {Haiman}}{2000}]{cen2000}
{Cen} R.,  {Haiman} Z.,  2000, \apjl, 542, L75

\bibitem[\protect\citeauthoryear{{Dietrich}, {Hamann}, {Shields}, {Constantin},
  {Heidt}, {J{\"a}ger}, {Vestergaard} \& {Wagner}}{{Dietrich}
  et~al.}{2003}]{dietrich2003}
{Dietrich} M.,  {Hamann} F.,  {Shields} J.~C.,  {Constantin} A.,  {Heidt} J.,
  {J{\"a}ger} K.,  {Vestergaard} M.,    {Wagner} S.~J.,  2003, \apj, 589, 722

\bibitem[\protect\citeauthoryear{{Fan}}{{Fan}}{2006}]{fanrev2006}
{Fan} X.,  2006, New Astronomy Review, 50, 665

\bibitem[\protect\citeauthoryear{{Fan}, {Strauss}, {Becker}, {White}, {Gunn},
  {Knapp}, {Richards}, {Schneider}, {Brinkmann} \& {Fukugita}}{{Fan}
  et~al.}{2006}]{fan2006}
{Fan} X.,  {Strauss} M.~A.,  {Becker} R.~H.,  {White} R.~L.,  {Gunn} J.~E.,
  {Knapp} G.~R.,  {Richards} G.~T.,  {Schneider} D.~P.,  {Brinkmann} J.,
  {Fukugita} M.,  2006, \aj, 132, 117

\bibitem[\protect\citeauthoryear{{Fan}, {Strauss}, {Richards}, {Hennawi},
  {Becker}, {White}, {Diamond-Stanic} \& {Donley}}{{Fan}
  et~al.}{2006}]{fan2006b}
{Fan} X.,  {Strauss} M.~A.,  {Richards} G.~T.,  {Hennawi} J.~F.,  {Becker}
  R.~H.,  {White} R.~L.,  {Diamond-Stanic} A.~M.,    {Donley} J.~L. e.~a.,
  2006, \aj, 131, 1203

\bibitem[\protect\citeauthoryear{{Furlanetto} \& {Oh}}{{Furlanetto} \&
  {Oh}}{2005}]{FurlanettoOh05}
{Furlanetto} S.~R.,  {Oh} S.~P.,  2005, \mnras, 363, 1031

\bibitem[\protect\citeauthoryear{{Furlanetto}, {Zaldarriaga} \&
  {Hernquist}}{{Furlanetto} et~al.}{2004}]{furlanetto2004}
{Furlanetto} S.~R.,  {Zaldarriaga} M.,    {Hernquist} L.,  2004, \apj, 613, 1

\bibitem[\protect\citeauthoryear{{Haiman}}{{Haiman}}{2002}]{haiman2002}
{Haiman} Z.,  2002, \apjl, 576, L1

\bibitem[\protect\citeauthoryear{{Hamann} \& {Ferland}}{{Hamann} \&
  {Ferland}}{1993}]{haman1993}
{Hamann} F.,  {Ferland} G.,  1993, \apj, 418, 11

\bibitem[\protect\citeauthoryear{{Jiang}, {Fan}, {Annis}, {Becker}, {White},
  {Chiu}, {Lin}, {Lupton}, {Richards}, {Strauss}, {Jester} \&
  {Schneider}}{{Jiang} et~al.}{2008}]{jiang2008}
{Jiang} L.,  {Fan} X.,  {Annis} J.,  {Becker} R.~H.,  {White} R.~L.,  {Chiu}
  K.,  {Lin} H.,  {Lupton} R.~H.,  {Richards} G.~T.,  {Strauss} M.~A.,
  {Jester} S.,    {Schneider} D.~P.,  2008, \aj, 135, 1057

\bibitem[\protect\citeauthoryear{{Jiang}, {Fan}, {Brandt}, {Carilli}, {Egami},
  {Hines}, {Kurk}, {Richards}, {Shen}, {Strauss}, {Vestergaard} \&
  {Walter}}{{Jiang} et~al.}{2010}]{jiang2010}
{Jiang} L.,  {Fan} X.,  {Brandt} W.~N.,  {Carilli} C.~L.,  {Egami} E.,  {Hines}
  D.~C.,  {Kurk} J.~D.,  {Richards} G.~T.,  {Shen} Y.,  {Strauss} M.~A.,
  {Vestergaard} M.,    {Walter} F.,  2010, \nat, 464, 380

\bibitem[\protect\citeauthoryear{{Jiang}, {Fan}, {Vestergaard}, {Kurk},
  {Walter}, {Kelly} \& {Strauss}}{{Jiang} et~al.}{2007}]{jiang2007}
{Jiang} L.,  {Fan} X.,  {Vestergaard} M.,  {Kurk} J.~D.,  {Walter} F.,  {Kelly}
  B.~C.,    {Strauss} M.~A.,  2007, \aj, 134, 1150

\bibitem[\protect\citeauthoryear{{Komatsu}, {Dunkley}, {Nolta}, {Bennett},
  {Gold}, {Hinshaw}, {Jarosik}, {Larson}, {Limon}, {Page}, {Spergel},
  {Halpern}, {Hill}, {Kogut}, {Meyer}, {Tucker}, {Weiland}, {Wollack} \&
  {Wright}}{{Komatsu} et~al.}{2009}]{komatsu2009}
{Komatsu} E.,  {Dunkley} J.,  {Nolta} M.~R.,  {Bennett} C.~L.,  {Gold} B.,
  {Hinshaw} G.,  {Jarosik} N.,  {Larson} D.,  {Limon} M.,  {Page} L.,
  {Spergel} D.~N.,  {Halpern} M.,  {Hill} R.~S.,  {Kogut} A.,  {Meyer} S.~S.,
  {Tucker} G.~S.,  {Weiland} J.~L.,  {Wollack} E.,    {Wright} E.~L.,  2009,
  \apjs, 180, 330

\bibitem[\protect\citeauthoryear{{Kramer} \& {Haiman}}{{Kramer} \&
  {Haiman}}{2008}]{kramer2008}
{Kramer} R.~H.,  {Haiman} Z.,  2008, \mnras, 385, 1561

\bibitem[\protect\citeauthoryear{{Kramer} \& {Haiman}}{{Kramer} \&
  {Haiman}}{2009}]{kramer2009}
{Kramer} R.~H.,  {Haiman} Z.,  2009, \mnras, 400, 1493

\bibitem[\protect\citeauthoryear{{Kurk}, {Walter}, {Fan}, {Jiang}, {Riechers},
  {Rix}, {Pentericci}, {Strauss}, {Carilli} \& {Wagner}}{{Kurk}
  et~al.}{2007}]{kurk2007}
{Kurk} J.~D.,  {Walter} F.,  {Fan} X.,  {Jiang} L.,  {Riechers} D.~A.,  {Rix}
  H.,  {Pentericci} L.,  {Strauss} M.~A.,  {Carilli} C.,    {Wagner} S.,  2007,
  \apj, 669, 32

\bibitem[\protect\citeauthoryear{{Lidz}, {McQuinn}, {Zaldarriaga}, {Hernquist}
  \& {Dutta}}{{Lidz} et~al.}{2007}]{lidz2007}
{Lidz} A.,  {McQuinn} M.,  {Zaldarriaga} M.,  {Hernquist} L.,    {Dutta} S.,
  2007, \apj, 670, 39

\bibitem[\protect\citeauthoryear{{Maselli}, {Ferrara} \& {Gallerani}}{{Maselli}
  et~al.}{2009}]{maselli2009}
{Maselli} A.,  {Ferrara} A.,    {Gallerani} S.,  2009, \mnras, 395, 1925

\bibitem[\protect\citeauthoryear{{Maselli}, {Gallerani}, {Ferrara} \&
  {Choudhury}}{{Maselli} et~al.}{2007}]{maselli2007}
{Maselli} A.,  {Gallerani} S.,  {Ferrara} A.,    {Choudhury} T.~R.,  2007,
  \mnras, 376, L34

\bibitem[\protect\citeauthoryear{{Mesinger}, {Haiman} \& {Cen}}{{Mesinger}
  et~al.}{2004}]{mesinger2004}
{Mesinger} A.,  {Haiman} Z.,    {Cen} R.,  2004, \apj, 613, 23

\bibitem[\protect\citeauthoryear{{Pentericci}, {Rix}, {Prada}, {Fan},
  {Strauss}, {Schneider}, {Grebel}, {Harbeck}, {Brinkmann} \&
  {Narayanan}}{{Pentericci} et~al.}{2003}]{pentericci2003}
{Pentericci} L.,  {Rix} H.,  {Prada} F.,  {Fan} X.,  {Strauss} M.~A.,
  {Schneider} D.~P.,  {Grebel} E.~K.,  {Harbeck} D.,  {Brinkmann} J.,
  {Narayanan} V.~K.,  2003, \aap, 410, 75

\bibitem[\protect\citeauthoryear{{Songaila}}{{Songaila}}{2004}]{songaila2004}
{Songaila} A.,  2004, \aj, 127, 2598

\bibitem[\protect\citeauthoryear{{Srbinovsky} \& {Wyithe}}{{Srbinovsky} \&
  {Wyithe}}{2007}]{srbinovsky2007}
{Srbinovsky} J.~A.,  {Wyithe} J.~S.~B.,  2007, \mnras, 374, 627

\bibitem[\protect\citeauthoryear{{Strateva}, {Brandt}, {Schneider}, {Vanden
  Berk} \& {Vignali}}{{Strateva} et~al.}{2005}]{strateva2005}
{Strateva} I.~V.,  {Brandt} W.~N.,  {Schneider} D.~P.,  {Vanden Berk} D.~G.,
  {Vignali} C.,  2005, \aj, 130, 387

\bibitem[\protect\citeauthoryear{{Telfer}, {Zheng}, {Kriss} \&
  {Davidsen}}{{Telfer} et~al.}{2002}]{telfer2002}
{Telfer} R.~C.,  {Zheng} W.,  {Kriss} G.~A.,    {Davidsen} A.~F.,  2002, \apj,
  565, 773

\bibitem[\protect\citeauthoryear{{Vanden Berk}, {Richards}, {Bauer}, {Strauss},
  {Schneider}, {Heckman}, {York}, {Hall}, {Fan}, {Knapp} \& {et al.}}{{Vanden
  Berk} et~al.}{2001}]{vandenberk2001}
{Vanden Berk} D.~E.,  {Richards} G.~T.,  {Bauer} A.,  {Strauss} M.~A.,
  {Schneider} D.~P.,  {Heckman} T.~M.,  {York} D.~G.,  {Hall} P.~B.,  {Fan} X.,
   {Knapp} G.~R.,    {et al.} 2001, \aj, 122, 549

\bibitem[\protect\citeauthoryear{{Vignali}, {Brandt}, {Schneider}, {Anderson},
  {Fan}, {Gunn}, {Kaspi}, {Richards} \& {Strauss}}{{Vignali}
  et~al.}{2003}]{vignali2003}
{Vignali} C.,  {Brandt} W.~N.,  {Schneider} D.~P.,  {Anderson} S.~F.,  {Fan}
  X.,  {Gunn} J.~E.,  {Kaspi} S.,  {Richards} G.~T.,    {Strauss} M.~A.,  2003,
  \aj, 125, 2876

\bibitem[\protect\citeauthoryear{{Wandel} \& {Petrosian}}{{Wandel} \&
  {Petrosian}}{1988}]{wandel88}
{Wandel} A.,  {Petrosian} V.,  1988, \apjl, 329, L11

\bibitem[\protect\citeauthoryear{{Wang}, {Carilli}, {Neri}, {Riechers}, {Wagg},
  {Walter}, {Bertoldi}, {Menten}, {Omont}, {Cox} \& {Fan}}{{Wang}
  et~al.}{2010}]{wang2010}
{Wang} R.,  {Carilli} C.~L.,  {Neri} R.,  {Riechers} D.~A.,  {Wagg} J.,
  {Walter} F.,  {Bertoldi} F.,  {Menten} K.~M.,  {Omont} A.,  {Cox} P.,
  {Fan} X.,  2010, ArXiv e-prints

\bibitem[\protect\citeauthoryear{{White}, {Becker}, {Fan} \& {Strauss}}{{White}
  et~al.}{2003}]{white2003}
{White} R.~L.,  {Becker} R.~H.,  {Fan} X.,    {Strauss} M.~A.,  2003, \aj, 126,
  1

\bibitem[\protect\citeauthoryear{{Willott}, {Albert}, {Arzoumanian},
  {Bergeron}, {Crampton}, {Delorme}, {Hutchings}, {Omont}, {Reyle} \&
  {Schade}}{{Willott} et~al.}{2010}]{willott2010}
{Willott} C.~J.,  {Albert} L.,  {Arzoumanian} D.,  {Bergeron} J.,  {Crampton}
  D.,  {Delorme} P.,  {Hutchings} J.~B.,  {Omont} A.,  {Reyle} C.,    {Schade}
  D.,  2010, ArXiv e-prints

\bibitem[\protect\citeauthoryear{{Willott}, {Delorme}, {Omont}, {Bergeron},
  {Delfosse}, {Forveille}, {Albert}, {Reyl{\'e}}, {Hill}, {Gully-Santiago},
  {Vinten}, {Crampton}, {Hutchings}, {Schade}, {Simard}, {Sawicki}, {Beelen} \&
  {Cox}}{{Willott} et~al.}{2007}]{willott2007}
{Willott} C.~J.,  {Delorme} P.,  {Omont} A.,  {Bergeron} J.,  {Delfosse} X.,
  {Forveille} T.,  {Albert} L.,  {Reyl{\'e}} C.,  {Hill} G.~J.,
  {Gully-Santiago} M.,  {Vinten} P.,  {Crampton} D.,  {Hutchings} J.~B.,
  {Schade} D.,  {Simard} L.,  {Sawicki} M.,  {Beelen} A.,    {Cox} P.,  2007,
  \aj, 134, 2435

\bibitem[\protect\citeauthoryear{{Willott}, {Delorme}, {Reyl{\'e}}, {Albert},
  {Bergeron}, {Crampton}, {Delfosse}, {Forveille}, {Hutchings}, {McLure},
  {Omont} \& {Schade}}{{Willott} et~al.}{2009}]{willott2009}
{Willott} C.~J.,  {Delorme} P.,  {Reyl{\'e}} C.,  {Albert} L.,  {Bergeron} J.,
  {Crampton} D.,  {Delfosse} X.,  {Forveille} T.,  {Hutchings} J.~B.,  {McLure}
  R.~J.,  {Omont} A.,    {Schade} D.,  2009, \aj, 137, 3541

\bibitem[\protect\citeauthoryear{{Wyithe}, {Bolton} \& {Haehnelt}}{{Wyithe}
  et~al.}{2008}]{wyithe2008}
{Wyithe} J.~S.~B.,  {Bolton} J.~S.,    {Haehnelt} M.~G.,  2008, \mnras, 383,
  691

\bibitem[\protect\citeauthoryear{{Wyithe} \& {Loeb}}{{Wyithe} \&
  {Loeb}}{2004}]{wyithe2004}
{Wyithe} J.~S.~B.,  {Loeb} A.,  2004, \nat, 427, 815

\bibitem[\protect\citeauthoryear{{Wyithe}, {Loeb} \& {Carilli}}{{Wyithe}
  et~al.}{2005}]{wyithe2005}
{Wyithe} J.~S.~B.,  {Loeb} A.,    {Carilli} C.,  2005, \apj, 628, 575

\end{thebibliography}

\label{lastpage}

\end{document}